\documentclass[journal,10pt]{IEEEtran}

\pdfoutput=1
\usepackage{algorithmicx}
\usepackage[ruled]{algorithm}
\usepackage{algpseudocode}
\usepackage{easyReview}

\ifCLASSOPTIONcompsoc
\usepackage[nocompress]{cite}
\else
\usepackage{cite}
\fi

\usepackage{amsmath}
\usepackage{txfonts}
\usepackage{array}
\usepackage{subfig}
\usepackage{hyperref}
\usepackage[hyphenbreaks]{breakurl}

\begin{document}

\title{Deep Reinforcement Learning Based Resource Allocation for Cloud Native Wireless Network}


\author{Lin Wang, Jiasheng~Wu,
	Yue~Gao,\IEEEmembership{~Senior Member, IEEE}, Jingjing Zhang}

\maketitle

\begin{abstract}
Cloud native technology has revolutionized 5G beyond and 6G communication networks, offering unprecedented levels of operational automation, flexibility, and adaptability. However, the vast array of cloud native services and applications presents a new challenge in resource allocation for dynamic cloud computing environments. To tackle this challenge, we investigate a cloud native wireless architecture that employs container-based virtualization to enable flexible service deployment. We then study two representative use cases: network slicing and Multi-Access Edge Computing. To optimize resource allocation in these scenarios, we leverage deep reinforcement learning techniques and introduce two model-free algorithms capable of monitoring the network state and dynamically training allocation policies. We validate the effectiveness of our algorithms in a testbed developed using Free5gc. Our findings demonstrate significant improvements in network efficiency, underscoring the potential of our proposed techniques in unlocking the full potential of cloud native wireless networks.
\end{abstract}



\section{Introduction}\label{introduction}
In the last few decades, there has been significant progress in the standardization of 5G/6G technology. The International Telecommunication Union (ITU) proposed a roadmap and vision for 5G in 2015 \cite{RN11}, and the official completion of the third version of 5G was marked with the freezing of the R17 standard in June 2022. While research on 6G has already begun \cite{6GVision}, there are still several technical challenges that must be addressed before 6G can realize its potential in driving business efficiencies, creating new insights, and enabling data monetization. Cloud native technology is essential in unlocking the full revenue potential of 5G/6G applications.

Cloud native is a software approach for building, deploying, and running applications in modern and dynamic environments, including public, private, and hybrid clouds \cite{RN24}. This approach utilizes various technologies such as containers, service meshes, microservices, and immutable infrastructure\cite{RN1}. Cloud native simplifies application development and enables more elastic scalability of applications. With the increasing demand for cloud computing and the advancement in cloud native technology, it has become a popular deployment solution for large software systems.

The increasing complexity and demand for network intelligence brought about by the widespread adoption of 5G networks provide an ideal scenario for cloud native technology. The high resiliency, scalability, and flexible orchestration offered by cloud native architecture have prompted many enterprises to introduce or plan to introduce cloud native technologies to their 5G networks. For instance, AT$\&$T has already shifted its 5G mobile network to Azure, Microsoft's cloud platform \cite{RN4}, and Telefonica has collaborated with Ericsson to implement a cloud-based 5G network \cite{RN2}.

However, with the proliferation of smart devices, interactive services, and intelligent applications, it is becoming clear that the 5G system will struggle to handle the massive volume of mobile traffic in the future\cite{6Gsurvey}. This challenge is driving the development of 6G, where cloud native architecture can provide a solid foundation by deploying the Radio Access Network (RAN) in the cloud. While 5G cloud native aims to move networks to the cloud, 6G cloud native systems will be natively integrated into the cloud. To facilitate the migration of the 5G network to the cloud, 3GPP has introduced the Service-Based Architecture (SBA) of 5G Core (5GC), which adopts cloud native design principles. 

The SBA of 5GC is composed of interconnected Network Functions (NFs) that are self-contained, independent, and reusable. Among these NFs, the User Plane Function (UPF) can be deployed near the network edge, which has spurred interest in edge activities like Multi-access Edge Computing (MEC). By leveraging 5G capabilities, MEC can offer benefits such as ultra-reliable and low-latency communications. Another unique feature of the 5G network is network slicing, which slices a single physical network infrastructure into several logical virtual networks to cater to the various demands of future networks. However, ensuring better Quality of Service (QoS) requires dynamic allocation of various network resources, which presents a critical challenge for both MEC and network slicing.

Resource allocation in network slicing and MEC has been the subject of extensive research, with most studies focusing on modeling methods \cite{RN6, RN7, RN28, RN29}. Ren et al. \cite{RN6} proposed an allocation algorithm based on queuing theory by modeling user requirements. Halabian et al. \cite{RN7} formulated the problem as a convex optimization problem using a closed-formed expression of the utility function and solved it with well-known convex optimization methods. Both Zheng et al. \cite{RN28} and Liu et al. \cite{RN29} formulated the resource allocation problem in multi-user computing networks as a stochastic game, achieving Nash equilibria. However, obtaining an accurate model in real 5G network circumstances is challenging, making these methods difficult to implement in real environments. Although these researchers obtained optimal solutions in their proposed models, the effectiveness of the solutions heavily depends on the accuracy of the models, which is hard to guarantee given the complexity of a 5G network.

Due to the difficulty of accurately modeling the environment, some researchers have employed model-free methods to solve resource allocation problems \cite{RN8, RN9, RN10, RN20}. For instance, Tang et al. \cite{RN8} proposed an allocation algorithm based on the Maximum and Minimum Ant Colony Algorithm, which is a type of swarm intelligence algorithm. Liu et al. \cite{RN9} successfully learned the relationship between utility and allocation policy using Deep Reinforcement Learning (DRL), solving the problem with classical optimization methods. In the context of MEC systems, Prandhan et al. \cite{RN17} applied a supervised Deep Learning (DL) method to address the computation-offloading problem for IoT applications in a C-RAN, and Liu et al. \cite{RN18} used the Convolutional Neural Network (CNN) for data analysis in a real-time MEC system. Similarly, Zhao et al. \cite{RN19} proposed a mobile data offloading strategy using an LSTM-based DL model to predict the real-time traffic of small base stations. Based on DRL, Lu et al. \cite{RN20} presented a resource allocation algorithm to reduce energy consumption and increase the usage of a large-scale heterogeneous network. Among these research studies, DRL has gained increasing attention as one of the promising solutions for 6G resource allocation problems \cite{RN9, RN10}. DRL is an advanced technique that combines powerful environmental perception with robust decision-making capabilities. As a result, it is well-suited to solving complex resource allocation problems in challenging circumstances. Researchers have proposed algorithms that utilize DRL or other model-free techniques, such as swarm intelligence, to allocate resources efficiently without needing a known model of the environment.

However, these studies are not based on actual 5GC environments, and their proposed solutions are usually only validated in simulated environments. These environments are often constructed using approximate models such as queues or chains, which cannot fully capture the complex features of 5GC environments. In other words, there is a discrepancy between the simulated environments and the real 5GC environments, which may result in poor performance of the proposed solutions in actual 5G networks.

The focus of this paper is to investigate resource allocation problems in a cloud native wireless network. The main contributions of our research can be summarized as follows:
\begin{itemize}
	\item We propose two model-free algorithms, based on the cloud native wireless architecture under investigation. The first is a TD3-based algorithm designed to allocate resources among network slices, while the second is a DQN-based algorithm for allocating computation resources in MEC.
	\item To verify the effectiveness of our proposed algorithms, we implement a small-scale cloud native testbed using open-source projects, i.e.,  Free5gc and UERANSIM.
	\item We test these algorithms in both simulated and real-time experimental 5G environments, using them as an example of 6G deployment. We show the effectiveness of the proposed algorithms.
 
\end{itemize}

The rest of the paper is organized as follows. In Section \ref{overview}, we give a brief overview of the cloud native architecture, and introduce the resource allocation problems, especially for network slicing and MEC; to solve the problems, two DRL algorithms, i.e., DQN and TD3 are reviewed. In Section \ref{NS}, we study the resource allocation problem in a network slicing model and propose a TD3-based solution, followed by a MEC network model and a DQN-based solution. In Section \ref{implementation}, we build a test platform to validate the proposed algorithms. The experimental results and analysis are presented in Section \ref{results}. Finally, a conclusion is drawn in Section \ref{conclusion}. 



\section{System Overview} \label{overview}
In this section, we first give an overview of the cloud native wireless network architecture under investigation. We then present resource allocation issues for two representative enabling technologies, i.e., network slicing and MEC. Finally, to address the issues, various DRL techniques for different application scenarios are briefly reviewed.



\subsection{Cloud Native Architecture}
\begin{figure}[t!] 
  \centering
\includegraphics[width=1\columnwidth]{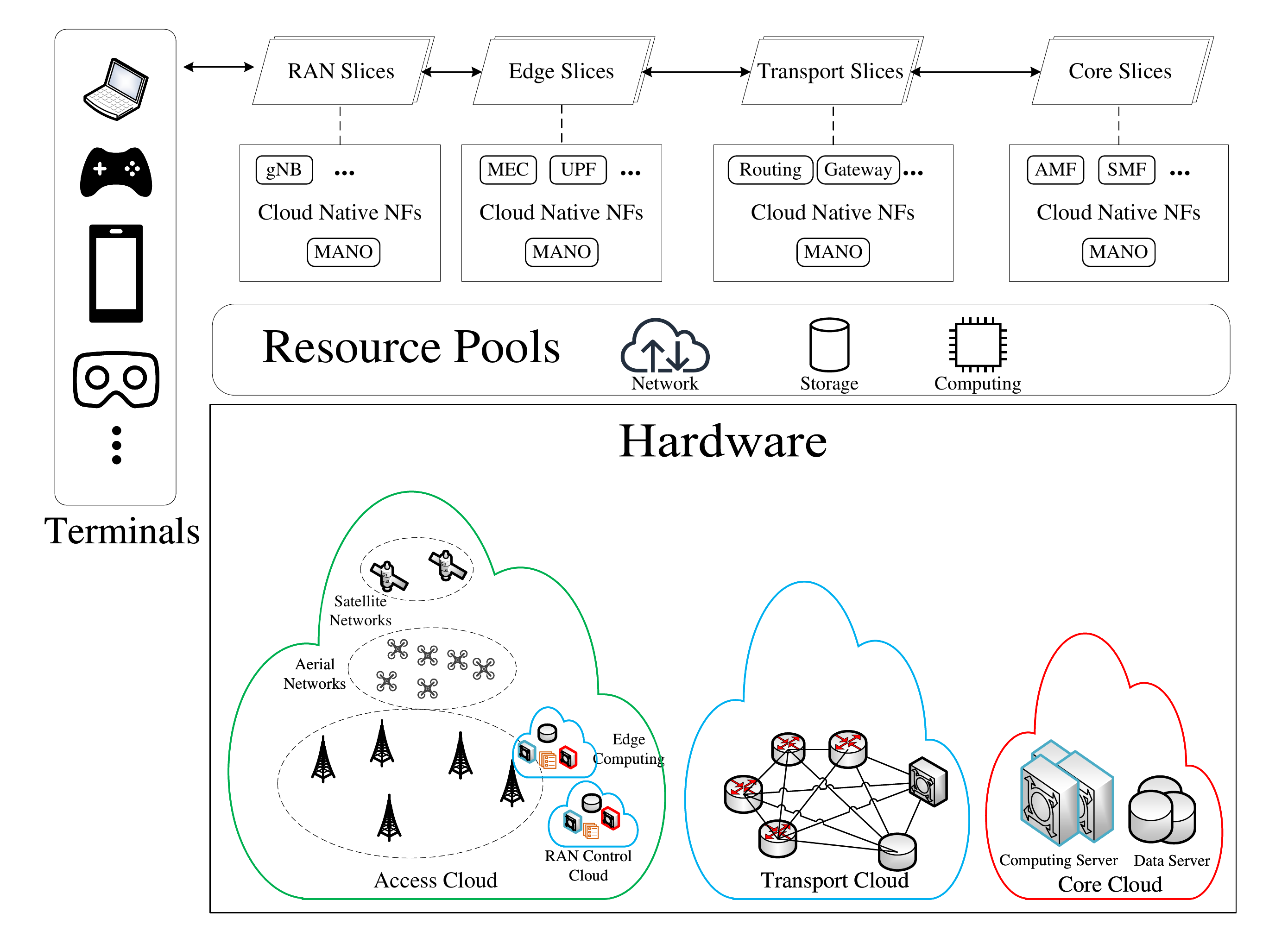}
\caption{Cloud native wireless architecture} 
\label{cloud-native system model}
\end{figure}


The architecture of a cloud native wireless network, which is currently being investigated for 5G beyond and 6G, is illustrated in Fig.~\ref{cloud-native system model}. It consists of a cloud native Core, a RAN, and a transport network that connects the RAN and the Core. By leveraging advanced technologies such as containerization, the cloud native system allows the creation of highly portable, isolated environments that can be quickly and easily deployed. This design enables the hardware components in a cloud native network to be separated into multiple virtual segments and shared among NFs. This, in turn, makes it possible to virtualize and dynamically allocate physical resources such as network, storage, and computing resources between different NFs, resulting in increased flexibility and efficiency of the system. 

There are several significant NFs listed in Fig.~\ref{cloud-native system model}, and we provide a brief introduction of them to aid in understanding. For example, the UPF is responsible for routing traffic to the desired destination. From a MEC system perspective, the UPF can be seen as a distributed and configurable data plane \cite{NF}. Additionally, the Access and Mobility Management Function (AMF) handles mobility-related procedures, while the Session Management Function (SMF) is responsible for selecting and controlling the UPFs. To ensure independent management, each NF is implemented in a cloud native way by introducing Management and Orchestration (MANO) to every single part. 

As a result, the cloud native architecture makes it possible to share infrastructure resources, enabling their dynamic allocation to meet various use cases. In this work, we focus on network slicing and MEC, with the objective of finding the optimal resource allocation strategies. 



Network slicing is a crucial technology that has been introduced to provide the necessary level of flexibility in 5G networks \cite{BK1}. By building network slices, network operators can divide the entire network into multiple virtual parts to meet different user requirements. However, before the introduction of the cloud native structure, NFs were the smallest unit for resource allocation in the Core. With the cloud native structure in the Core, it has become possible to allocate resources to an even smaller unit, where the virtualized resources (such as computation and communication abilities) of machines can be allocated dynamically. This feature provides network operators with the ability to design a dynamic resource allocation policy, maximizing the network's utility. By enabling the allocation of resources to smaller units, the cloud native structure enhances the efficiency of resource utilization, leading to cost savings, scalability, and increased flexibility for the network.


MEC is also a vital technology in the deployment of cloud native wireless networks. By leveraging the computing capabilities and applications at the network's edge, such as at the base station or access point, MEC provides distributed computing capabilities that enable operators to deliver low-latency, high-bandwidth, and localized services to end-users. While MEC greatly strengthens the serviceability of edge networks, it also requires appropriate resource allocation strategies. Since edge resources are distributed and scattered in the edge network, it is a waste of resources if available scattered ones cannot be efficiently utilized\cite{MEC_allocation}. However, with an effective resource allocation strategy, edge resources can be distributed properly, enabling faster task processing and improving edge system utility.

\subsection{Deep Reinforcement Learning Preliminaries}
DRL has been widely adopted to solve resource allocation or other decision-making problems, as mentioned in Section \ref{introduction}. Here we briefly provide some preliminaries on DRL algorithms, mainly focusing on DQN and TD3 algorithms that are related to our work.


\subsubsection{DQN}


DQN, or Deep Q-Network, is an extension of classic Q-learning that utilizes value functions \cite{RN13}. To approximate the Q-value, DQN employs a deep neural network with adjustable parameters $\boldsymbol{\theta}$. The algorithm enhances performance by using a replay buffer and a target network. The target network has parameters $\boldsymbol{\theta'}$, which are periodically copied from $\boldsymbol{\theta}$ and kept constant for a specified number of iterations, thereby improving the network's stability. It is important to note that DQN is appropriate for decision-making problems that involve discrete action spaces.

\subsubsection{TD3}
TD3, or Twin Delayed Deep Deterministic Policy Gradient\cite{RN15}, is a combination of continuous Double Deep Q-Learning, Policy Gradient, and Actor-Critic algorithms, making it capable of handling environments with continuous action spaces. TD3 addresses the issue of Q-value overestimation by using two identical Critic networks to estimate the Q-value, then selecting the smaller estimation. Moreover, TD3 employs action noise and learns with a non-deterministic policy, resulting in better robustness and applicability to more complex scenarios.

Since TD3 is more suitable for the continuous action space while DQN performs better with a discrete one, we adopt TD3 algorithm for the resource allocation among multiple network slices and DQN is used to find the optimal task offloading strategy in an MEC network. 



\section{Resource Allocation for Network Slicing} \label{NS}
 In this section, we first present a network slicing model based on the cloud native network. Then, a TD3-based algorithm is proposed to find the optimal resource allocation strategy among network slices.

\subsection{Network Slicing Model}
The scenario depicted in Fig.~\ref{cloud-native system model} involves multiple network slices that share computation and communication resources stored in the resource pools. We virtualize the physical resources in the Core and allocate them to different network slices, i.e., NFs in the network slices. The UPF is responsible for forwarding all users' data, with its bandwidth being the primary bottleneck of the network. Thus, we examine the bandwidth allocation problem for UPFs.

We assume that the overall bandwidth $B$ of the network is some constant and there are $I$ slices in total in the network. To elaborate on the slice allocation strategy, we define the vector $\boldsymbol{k} \triangleq\left[k_{i}, i=1,\cdots,I \right]$, where each element $k_i$ denotes the bandwidth that slice $i$ is allocated. Therefore, we have the constraint
\begin{equation}
	\sum_{i=1}^{I} k_{i} \leq B.
\end{equation}
To satisfy the Service Level Agreement (SLA) of the slices, we set an upper limit $k_{\max, i}$ and a lower limit $k_{\min , i}$ for the bandwidth each slice can get, yielding the following inequality
\begin{equation}
	k_{\min , i} \leq k_{i} \leq  k_{\max , i}, i=1,\cdots,I .
\end{equation}

Denote the vector $\boldsymbol{c} \triangleq$ $\{c_{i}, i=1,\cdots,I \}$ as the score vector in the network, where each $c_i$ represents the score of slice $i$. $\boldsymbol{c}$ is generally decided by the performance of the network, such as the average latency, the average transport speed, and the user demand satisfaction. These factors are crucial to user experience and are heavily influenced by our network slicing strategy. 
%
To evaluate the situation of the whole network system, we define the utility function $U$ given as
\begin{equation}
	U \triangleq \prod_{i=1}^{I}  c_{i}. \label{utility}
\end{equation}
The target of our resource allocation decisions is to optimize the user's experience, i.e., to find an allocation strategy $\boldsymbol{k}$ that can maximize the network's utility function $U$ with the problem formulated as
\begin{align}
	\mathcal{P}_{1}: &~\underset{\boldsymbol{k}}{\operatorname{argmax}} ~~~ U, 
            \notag   \\
	&\text { s.t. } \text { (1) (2) .}  \label{target}
\end{align}

\begin{algorithm}[t!]
	\caption{TD3-based resource allocation algorithm}
	\begin{algorithmic}[1]
  \State $\textbf{Initialization}$: 
		\State\hspace{\algorithmicindent} Initialize \textbf{\textit{critic network}} $Q_{\theta_{1}},Q_{\theta_{2}}$, \textbf{\textit{actor network}} $\pi_{\phi}$, \\ \ \ \ \ \  where $\theta_{1}, \theta_{2}, \phi$ are the parameters
		\State\hspace{\algorithmicindent} Initialize \textit{\textbf{Target Network}} $Q_{\theta_{1}^{\prime}}, Q_{\theta_{2}^{\prime}}, \pi_{\phi^{\prime}},\theta_{1}^{\prime} \leftarrow \theta_{1},$ \State\hspace{\algorithmicindent} $\theta_{2}^{\prime} \leftarrow \theta_{2}$,
		$\phi^{\prime} \leftarrow \phi$, learning rate $\tau$ 
		\State\hspace{\algorithmicindent} Initialize \textbf{\textit{Replay Buffer}} $RB$
         \State\hspace{\algorithmicindent} Maximum training episodes $T$, maximum exploration\\ \ \ \ \ \ episodes $T_1$ and noise variance $\sigma$
        \For {$t=1,\ldots,T$}
         \If {$t\leq T_1$}
        \State $\textbf{Exploration}$
        \State take a random action $\boldsymbol{a}$
        \Else
        \State $\textbf{Exploration \& Exploitation}$
		\State action $\boldsymbol{a} \sim \pi_{\phi}(s)+\epsilon$, with $\epsilon \sim N(0, \sigma)$
    \EndIf
		\State calculate allocation $\boldsymbol{k}$ according $\boldsymbol{a}$
		\State execute the allocation $\boldsymbol{k}$
		\State observe reward $r$ and new state $s^{\prime}$, record
		$\left(\boldsymbol{s}, \boldsymbol{a}, r, \boldsymbol{s}^{\prime}\right)$
		\State select mini-batch from $RB$ and update the networks:
		\State \begin{equation*}
			\begin{aligned}
				&\widetilde{a} \leftarrow \pi_{\phi^{\prime}}\left(\boldsymbol{s}^{\prime}\right)+\epsilon, y \leftarrow r+\gamma \min _{\mathrm{i}=1,2} Q_{\theta_{i}^{\prime}}\left(\boldsymbol{s}^{\prime}, \widetilde{\boldsymbol{a}}\right) \\
				&\theta_{i} \leftarrow \operatorname{argmin}_{\theta_{i}} N^{-1} \sum\left(y-Q_{\theta_{i}}(\boldsymbol{s}, \boldsymbol{a})\right)^{2}
			\end{aligned}
		\end{equation*}
        \If{$t\bmod d \equiv 0$}
        \State update $\phi$ by the deterministic policy gradient 
        \State update $\theta_{1}^{\prime}, \theta_{2}^{\prime}, \phi^{\prime}$: 
        \State \begin{equation*}
			\begin{aligned}
				&\theta_{i}^{\prime} \leftarrow \tau\theta_{i} + (1-\tau)\theta_{i}^{\prime}\\
				&\phi^{\prime} \leftarrow \tau\phi+ (1-\tau)\phi^{\prime}
			\end{aligned}
		\end{equation*}
        \EndIf

\EndFor
\State \textbf{return} $\pi_{\phi}$
	\end{algorithmic}
\end{algorithm}

\subsection{TD3-Base Algorithm Design} 
In a real environment, it's difficult to build an accurate model of the whole network. Hence, we can not rely on a closed-form utility function to solve the problem $\mathcal{P}_{1}$. 
To solve this issue, we use the Markov decision process (MDP) to formulate our resource allocation problem, which is a decision-making framework for modeling dynamic systems in stochastic and uncertain environments. 

A MDP can be characterized by a tuple $\langle S, A, R, P, \gamma\rangle$. Here, $S$ refers to the set of states of all the network slices, and $A$ represents the set of actions that the slices can execute. The reward function $R$ determines what reward each slice will get when taking an action in a particular state. The state transition function $P$ defines the probability of moving from one state to another when a particular action is taken. Finally, $\gamma$ determines how the reward is discounted over time. In detail, the first three items $S, A, R$ are defined below.

\subsubsection{\textbf{State}} $\boldsymbol{s} \triangleq \{s_{i},  i=1,\cdots, I\}, \boldsymbol{s} \in S$ contains the state of each slice in the network. 
As the overall states of the real environment are often not fully observable, here we select several parameters that are easy to observe and can comprehensively reflect the system state. The state of slice $i$ at time step $t$ is given by 
\begin{equation}
	s_{\boldsymbol{i}}^t = (o_i^t,l_i^t,d_i^t),
\end{equation}
where $o_i^t = k_i^{t-1}$ is the resource that slice $i$ gets in the previous time step $t-1$, $l_i^t$ is the average latency of all services in slice $i$ and $d_i^t$ is the average traffic of slice $i$. In order to measure the performance of the algorithm, we normalize all the components.

\subsubsection{\textbf{Action}}
We define the action vector $\boldsymbol{a} \triangleq\{a_{i},  i=1,\cdots,I\}$ of the set $A$, where $a_{i} \in$ $[-1,1]$ is the action for slice $i$ that determines how much resource slice $i$ can obtain. More precisely, $ a_{i}=-1 $ indicates that slice $i$ is allocated the least amount of bandwidth resource, i.e., we have $k_i=k_{min,i}$. On the contrary, $a_{i}=1$ represents that we have $k_i=k_{max,i}$. With the action vector $\boldsymbol{a}$, we can calculate the bandwidth resource that slice $i$ obtains, given as 
\begin{equation} 
    k_i = \min\Bigg\{k_{max,i},k_{min,i}+\Big(n-\sum_{j \in I} k_{min,j}\Big) * \frac{(a_i+1)}{\sum_{j \in I}(a_j+1)}\Bigg\}. \label{allocate}
\end{equation}

\subsubsection{\textbf{Reward}}
We define the network's utility $U$ as the reward $r$ obtained when taking the action $\boldsymbol{a}$ in the state $\boldsymbol{s}$, given as 
\begin{equation}
	r(\boldsymbol{s}, \boldsymbol{a})=U.
\end{equation}

Our object is to find the optimal policy $\pi^{*}$ that provides the action able to maximize the discounted cumulative reward within a given $T$ time step, which is calculated as
\begin{equation}
    \mathop{\min} \sum_{t=1}^T \gamma^{T-t} r^t.
\end{equation}

We use the TD3 algorithm to train neural networks as the optimal policy $\pi^{*}$, with a two-layer fully-connected neural network as the critic and a three-layer fully-connected neural network as the actor, each with 256 neurons. We also create a Target Network with the same structure for each neural network. The training process is divided into two phases, an exploration phase ($T_1$ episodes) and an exploration-exploitation phase ($T-T_1$ episodes). During exploration, the algorithm takes random actions, while during exploitation, it selects the action $\boldsymbol{a}$ based on the current state $\boldsymbol{s}$, actor-network $\pi_{\phi}$, and random noise. The algorithm then calculates and executes the allocation $\boldsymbol{k}$ using Eq. \ref{allocate}, receives immediate rewards $\boldsymbol{r}$, and updates the Replay Buffer and strategy. At each step, the algorithm selects several records from the Replay Buffer to train the network and updates the weight of the Target Networks to follow the changes of the online networks $Q_{\theta_{1}}, Q_{\theta_{2}}$ and $\pi_{\phi}$. The pseudo-code of the proposed TD3-based resource allocation algorithm is presented as Alg. $1$.


\section{Resource Allocation for MEC Network}\label{MEC}
We proceed to address the computing resource allocation problem in an MEC network. To this end, a DQN-based algorithm is designed to tackle the allocation optimization problem.

\subsection{MEC Model}

\begin{figure}[t!] 
  \centering
\includegraphics[width=0.85\columnwidth]{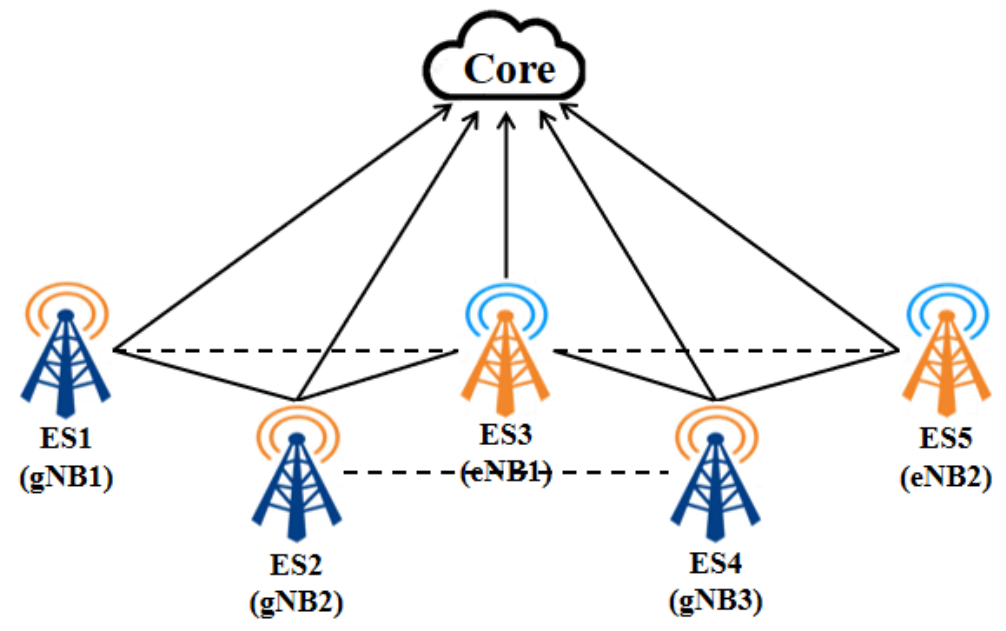}
\caption{An edge computing network with $I$ Edge Servers and the Core.} 
\label{mec system model}
\end{figure}

A more specific MEC network model under the architecture in Fig.~\ref{cloud-native system model} is investigated. As shown in Fig.~\ref{mec system model}, we consider an edge computing network in which a set $\mathcal{I}=\{1,2,\ldots,I\}$ of $I$ Edge Servers (ES) are connected to the Core; each ES is a RAN node, either a next Generation NodeB (gNB, i.e., a 5G base station) or an enhanced NodeB (eNB, i.e., an enhanced 4G base station). It is assumed that each ES $i$ is able to communicate with $N_i$ neighboring ESs, denoted by the set $\mathcal{N}_i\in \mathcal{I}\backslash\{i\} $, within some communication radius. Take ES $3$ as an example, it can communicate with 4 neighboring ESs, i.e., we have $N_3=4$ and $\mathcal{N}_3=\{1,2,4,5\}$. Moreover, due to the variety of the RAN nodes, the ESs have different but limited computation capacities, denoted by the set $\mathcal{C}=\{C_1,C_2,\ldots,C_I\}$. 

Consider a time-slotted system with $T$ slots in total, indexed by the set $\mathcal{T}=\{1,2,\ldots, T\}$. At each time slot $t$, each ES $i$ receives computation tasks $\mathcal{L}_i^t$ of size $S_i^t$ from one or multiple end users. Due to limited computation capacity, ES $i$ can process tasks of size $\hat{S}_i=\tau C_i/v$ at most within constrained computing latency $\tau$. For each ES $i$, if the received tasks have size $S_i^t < \hat{S}_i$ at time slot $t$, it would process them all locally. Otherwise, the tasks $\mathcal{L}_i^t$ are assumed to be split into two parts. The first part $\hat{\mathcal{L}}_i^t$ with size $\hat{S}_i$ is executed locally by ES $i$. The size of the second part $\bar{\mathcal{L}}_i^t$ 
is hence given as
 \begin{equation}
    \bar{S}_i^t = S_i^t - \hat{S}_i = S_i^t - \tau C_i/v,
\end{equation}
here $v$ represents the number of CPU cycles needed to process 1 bit data.
 
Each ES $i$ can process $\bar{\mathcal{L}}_i^t$ in two possible ways: 1) offload $\bar{\mathcal{L}}_i^t$ to some neighboring ES which has enough available computing resources; 2) forward $\bar{\mathcal{L}}_i^t$ to the Core for remote computation if none of the neighboring ESs has sufficient extra computing resources. As a result, each ES $i$ may receive multiple requests from neighbors. In order to increase resource efficiency, it would accept the one which could make the most of its available computing resources in the first place. Since it is assumed that the Core has an incredibly massive computation capacity, its computing latency can be ignored, thus effectively reducing total task processing latency. However, transmitting $\bar{\mathcal{L}}_i^t$ to a neighboring ES will increase the resource utilization efficiency of the edge network.


Overall, at each time slot $t$, for ES $i$, the received task $\mathcal{L}_i^t$ can be carried out in the following three ways, represented by the flag $f_i^t\in\{-1,0,1\}$. To elaborate, we denote the processing latency of task $\mathcal{L}_i^t$ by $L_i^t$. 

1) Compute $\mathcal{L}_i^t$ locally only with $f_i^t=1$: Given the computation capacity $C_i$, the processing latency $L_i^t$ is equal to the computing latency, given as
\begin{equation}
    L_i^t = v S_i^t/C_i.
\end{equation}

2) Forward $\bar{\mathcal{L}}_i^t$ to the Core with $f_i^t=0$: In this case, the total task processing latency consists of the computation latency at ES $i$ and the transmission delay from ES $i$ to the Core since the corresponding computation time can be ignored. Hence, we have
\begin{equation}
    L_i^t =\tau + \bar{S}_i^t/R_c,
\end{equation}
where each ES is assumed to have the same link capacity $R_c$ bits to the Core.

3) Offload $\bar{\mathcal{L}}_i^t$ to the neighboring ES $j$ with $f_i^t=-1$: ES $i$ is not able to process the whole task $\mathcal{L}_i$ within time limit $\tau$ and hence transmits $\bar{\mathcal{L}}_i$ to ES $j$ having sufficient resources. As a result, the total task processing latency is calculated as
\begin{eqnarray}
    L_i^t=\tau+\bar{S}_i^t/R_{i,j} + v\bar{S}_i^t/C_j,
\end{eqnarray}
where $R_{i,j}$ is the data capacity between ES $i$ and $j$.

We define the maximum latency $L(t)$ of all the tasks as a measure of the performance, given as
\begin{equation}
    L(t) = \max_{i\in \mathcal{I}} \{L_i^t\}.
\end{equation}


\subsection{DQN-Based Algorithm Design}
In the first place, we transform the proposed multi-server resource allocation case to a Markov decision process, still denoted by the tuple $\langle S, A, R, P, \gamma\rangle$. Here, $S=\left\{ s_1, s_2, \ldots, s_I \right\}$ is the joint state set that includes the state of each ES; $A=\left\{ a_1, a_2, \ldots, a_I\right\}$ refers to the set of ESs' actions; $R=\left\{ r_1, r_2, \ldots, r_I\right\}$ contains the rewards observed by all the ESs; $P$ represents the state transition probability function and $\gamma$ is the discounted factor.

Therefore, we can formulate the considered task offloading and resource allocation problem as a Markov process game. The elaborations of the first three items in the tuple $\langle S, A, R, P, \gamma\rangle$ are given below.

\subsubsection{\textbf{Action}}  In our proposed system, at any time slot $t$, each ES $i$ receives task $\mathcal{L}_i^t$ of size $S_i^t$. If $S_i^t \leq \hat{S}_i^t$ ($f_i^t=1$), 
the task would be executed locally at once by ES $i$. Otherwise, $\mathcal{L}_i^t$ is going to be split into two parts as mentioned before. ES $i$ processes the first part $\hat{\mathcal{L}}_i^t$ locally; and for the second part $\bar{\mathcal{L}}_i^t$, ES $i$ would take an action to determine how to execute it. We designate $a_i^t$ as ES $i$'s action at time slot $t$, i.e.,
\begin{equation}
    a_i^t = \begin{cases}
     -1 & \text{if $f_i^t=0$} \\ j, j\in\mathcal{N}_i & \text{if $f_i^t=-1$.}
     \end{cases}
\end{equation}
To elaborate, $a_i^t=-1$ represents that ES $i$ would transmit $\bar{\mathcal{L}}_i^t$ to the Core while $a_i^t=j$ indicates that ES $i$ offloads $\bar{\mathcal{L}}_i^t$ to the neighboring ES $j$. 

Obviously, the action space here is discrete. All the $I$ ESs' actions constitute the joint action $\boldsymbol{A}^t=\left\{ a_1^t, a_2^t, \ldots, a_I^t \right\}$, and the whole action space is the set of all the possible joint actions, denoted by $\mathcal{A}=\left \{\boldsymbol{A_1}, \boldsymbol{A_2}, \ldots, \boldsymbol{A_N} \right \}$.

\subsubsection{\textbf{State}} The state vector $\boldsymbol{S}=\left \{s_1, s_2, \ldots, s_I\right \}$ contains the status of all the $I$ ESs in the edge computing network. Here, the state observed by ES $i$ at time slot $t$ is given by
\begin{equation}
    s_{i}^t=( d_i^t, o_i^t),
\end{equation}
where $d_{i}^t=L_i^{t-1}$ is the processing latency of the tasks ES $i$ received at the last time slot, and $o_i^t=a_i^{t-1}$ indicates the way of task offloading at the previous time slot. 




\subsubsection{\textbf{Reward}} 
As mentioned before, the reward set $\boldsymbol{R}^t=\left\{ r_1(s_1^t, a_1^t), r_2(s_2^t, a_2^t), \ldots, r_I(s_I^t, a_I^t)\right\}$ contains the rewards obtained by all $I$ ESs. The reward $r_i(s_i^t, a_i^t)$ of ES $i$ in state $s_{i}^t$ represents the immediate return by ES $i$ taking action $a_i^t$, given by
\begin{equation}
    r_i(s_i^t, a_i^t)=L_i^t,
\end{equation}
where $L_i^t$ is the processing latency of computation task $\mathcal{L}_i^t$. 
%


\begin{algorithm}[t!]
    \caption{DQN-based resource allocation algorithm\label{Algorithm 2}}
    \begin{algorithmic}[1]
        \State $\textbf{Initialization}$: 
        \State\hspace{\algorithmicindent} Initialize \textbf{\textit{Q Network}} $Q_{\theta}$ with random weights $\theta$
        \State\hspace{\algorithmicindent} Construct a \textit{\textbf{Target Network}} $Q_{\theta}^{\prime}$ with weights $\theta^{\prime}$
        \State\hspace{\algorithmicindent} Initialize the \textbf{\textit{Replay Buffer}}
        \State\hspace{\algorithmicindent} Maximum training episodes $T$ and maximum explo\\ \ \ \ \ \  -ration episodes $T_1$
        \State\hspace{\algorithmicindent} exploration-exploitation rate $\epsilon$
        \For {$t=1,\ldots,T$}
        \If {$t\leq T_1$}
        \State $\textbf{Exploration}$
        \State Take a random action $\boldsymbol{A}^t$
        \Else
        \State $\textbf{Exploration \& Exploitation}$
        \State Generate a random number $randnum$
        \If {$randnum < \epsilon$}
        \State Take a random action $\boldsymbol{A}^t$
        \Else
        \State Choose action $\boldsymbol{A}^t$ according to $\mathop{\mathrm{argmin}}\limits_{\boldsymbol{A}\in\mathcal{A}} Q(\boldsymbol{S}^t, \boldsymbol{A})$
        \EndIf
        \EndIf
        \State Execute task offloading and processing according to $\boldsymbol{A}^t$
        \State Observe reward $\boldsymbol{R}^t$ and new state $\boldsymbol{S}^{t+1}$ 
        \State Record $\left(\boldsymbol{S}^t, \boldsymbol{A}^t, \boldsymbol{R}^t, \boldsymbol{S}^{t+1}\right)$ in Replay Buffer
        \State Select mini-batch from Replay Buffer and update loss \\ \ \ \ \ function:
        \begin{equation*}
            loss \leftarrow [R_{t}+\gamma  Q_{\theta^{t+1}}(\boldsymbol{S}^{t+1}, \boldsymbol{A})-Q_{\theta^{t}}(\boldsymbol{S}^{t}, \boldsymbol{A}^t)]^{2}
        \end{equation*}  
        \State Update $Q$-value
        \If{$t\bmod d \equiv 0$}
        \State Update \textit{\textbf{Target Network}} $Q_{\theta}^{\prime}$, $\theta^{\prime} \leftarrow \theta$
        \EndIf
        \EndFor

    \end{algorithmic}
\end{algorithm}

Therefore, the objective of the proposed problem is to minimize the discounted cumulative total negative reward concerning all ESs over the given $T$ time slots, which can be formulated as
\begin{equation}
    \mathop{\min} \sum_{t=1}^T \gamma^{T-t} R^t,
\end{equation}
here parameter $R^t$ is the maximum reward among all the ESs at time slot $t$, given as $R^t=\max\{r_i(s_i^t, a_i^t)\}=\max\{L_i^t\}=L(t)$.

As the action space $\mathcal{A}$ is discrete, here we deploy DQN algorithm to train neural networks and then provide the optimal action-selecting strategy that is able to minimize the task processing latency at any time slot $t$, thus decreasing total system cost in the long term.

We define task offloading strategy as $\pi$, which is a mapping from the state space $\boldsymbol{S}$ to the action space $\mathcal{A}$. The state-action value $Q(\boldsymbol{S}^t, \boldsymbol{A}^t)$ represents the obtained reward when ESs taking joint action $\boldsymbol{A}^t$ in state $\boldsymbol{S}^t$ under strategy $\pi$ at time slot $t$, and we use it to select the optimal action that minimizes the $Q$-value in a given state, given by
\begin{equation}
    \boldsymbol{A}^t = \mathop{\mathrm{argmin}}\limits_{\boldsymbol{A}\in\mathcal{A}} Q(\boldsymbol{S}^t, \boldsymbol{A}).
\end{equation}

The pseudo-code of the proposed DQN-based resource allocation algorithm in the discrete space is presented as Alg .\ref{Algorithm 2}. We use a 3-layer fully-connected neural network with weight set $\theta$ as a Q Network and the same one as a Target Network. The whole training process including $T$ episodes is divided into two phases, i.e., exploration phase ($T_1$ episodes) and exploration-exploitation phase ($T-T_1$ episodes). When exploring, ESs would take random actions. While for exploitation, the ESs choose actions $\boldsymbol{A}^t$ in state $\boldsymbol{S}^t$ according to the current minimum state-action value $Q(\boldsymbol{S}^t, \boldsymbol{A})$. After executing tasks, ESs receive immediate rewards $\boldsymbol{R}^t$, and then the edge computing network enters into the next state $\boldsymbol{S}^{t+1}$. The network would record these variables in the replay buffer and update Q-value and the task-offloading strategy. After the given $d$ steps, we copy the weight set $\theta$ of Q Network to $\theta'$ of Target Network.

In our proposed algorithm, we adopt the $\epsilon$-greedy method as the action selection strategy to balance exploration and exploitation. During the exploration-exploitation phase, ES $i$ randomly chooses an action with a probability of $\epsilon\in [0,1]$, and selects the optimal action with the minimum known Q-value with a probability of $1-\epsilon$. A higher value of $\epsilon$ indicates a greater tendency towards exploration.

\section{System Implementation}\label{implementation}

\begin{figure}[t!]
	\centering
	\includegraphics[width=3.3in]{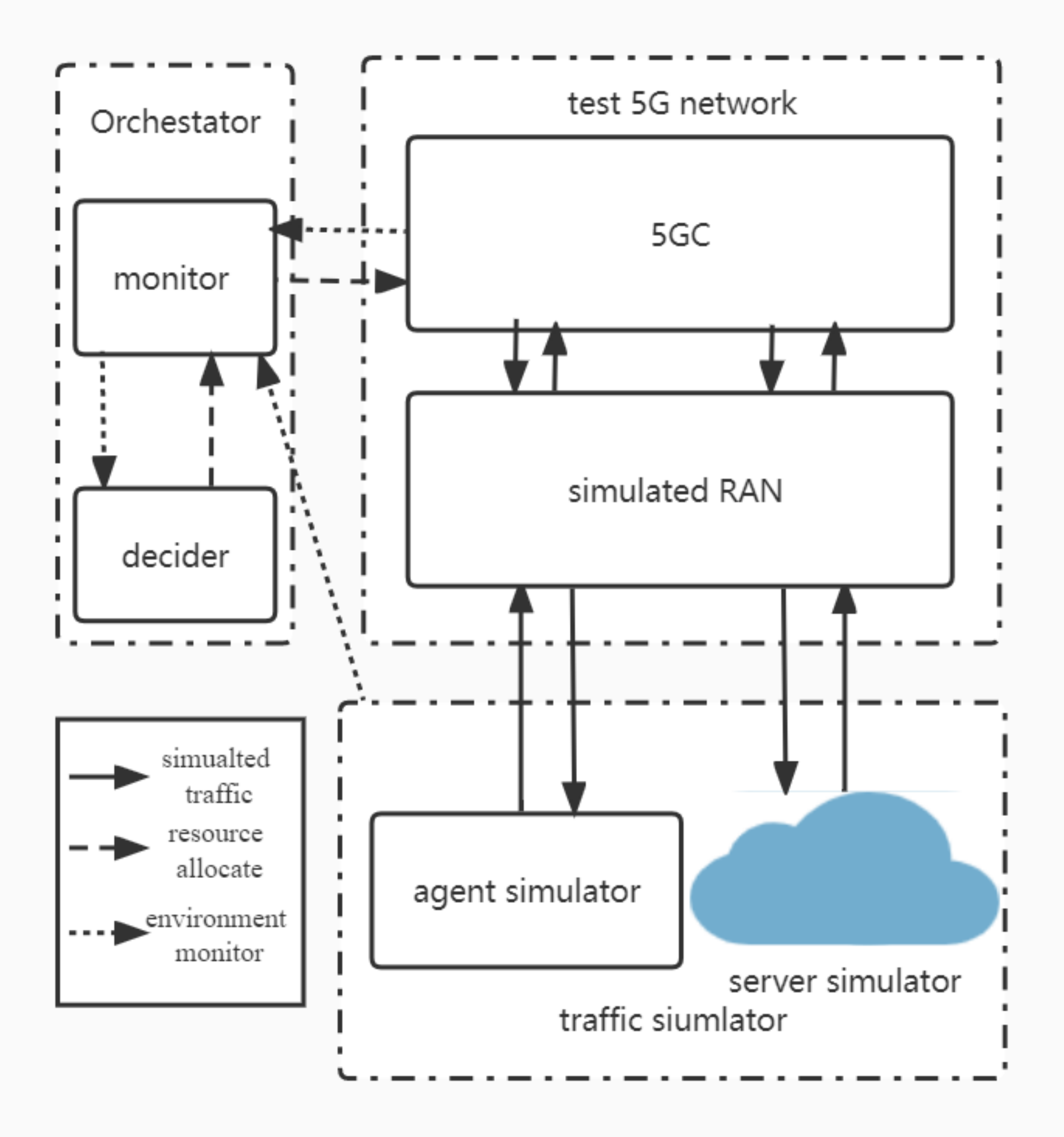}
	\caption{Structure of 5G resource allocation testbed.}
	\label{Fig2}
\end{figure}

We have designed a versatile experimental platform to accommodate a wide range of 5G scenarios, as illustrated in Fig.~\ref{Fig2}. Specifically, we made targeted modifications to two models within our study, i.e., network slicing and MEC. Our testbed is composed of a cloud native 5G network, an orchestrator, and a traffic simulator. The 5G network serves as the backbone of our platform, while the orchestrator monitors the entire network and allocates resources as needed. The traffic simulator generates realistic traffic in the network, allowing us to evaluate the performance of the environment. We proceed to delve into the details of each individual component of our testbed.

\subsection{Test 5G network}
The test 5G network has two parts: 5GC and simulated RAN. The 5GC is a cloud native one in which the NFs are deployed in independent containers and hardware resources can be allocated agilely among the NFs.


The 5GC of our test 5G network is based on the open-source project Free5gc \cite{free5gc}. The main objective of this project is to implement the 5GC defined in 3GPP Release 15 (R15) and beyond. Free5gc has successfully implemented the 5G architecture specified by the 3GPP R15 standard, including all necessary network elements such as UPF, AMF, and SMF. In particular, Free5gc supports network slicing and provides the required support for MEC on the UPF side, hence making it an ideal choice for our two different experimental scenarios. Moreover, Free5gc strictly adheres to the 3GPP standard, making the test 5G network built by Free5gc a 5G core network that conforms to a realistic environment. As Free5gc does not natively support cloud native deployment, we modified the software to deploy all network elements in separate containers using Docker \cite{docker}.

The simulated RAN is designed based on UERANSIM. UERANSIM \cite{ueransim} is the open-source state-of-the-art 5G User Equipment (UE) and RAN (gNodeB) implementation. In the network slicing case, we build a simulated gNodeB and several simulated UEs connected to the gNodeB simulator. Each simulated UE has a unique S-NSSAI to select its network slice. While as to the MEC case, we set up several simulated gNodeBs and UEs with the same number. Each UE is connected to all gNodeBs and these gNodeBs are allowed to communicate with each other.

\subsection{Traffic simulator}
We implement the traffic simulator using Golang 1.17 \cite{golang}. The simulator comprises a server simulator and an agent simulator. The agent simulator sends requests through a specific port and evaluates the performance of the proposed networks, while the server simulator handles all these requests. 

Using our traffic simulator, we simulate three different network services in the network slicing model, each with unique traffic characteristics. These services include video service, voice service, and online chat service. For video service, a large video file is divided into several small parts that are transferred sequentially, resulting in intermittent traffic. Voice service, on the other hand, utilizes Voice over Internet Protocol (VoIP) technology, where packets of the same size are sent at regular intervals, resulting in stable and periodic traffic. As for online chat service, we assume the traffic to be random due to the unpredictable nature of chat conversations. Overall, our traffic simulator allows us to simulate and evaluate the performance of each service within the network slicing model. The traffic characterization of these three services is shown in Table \ref{Table1}.

\begin{table}[!t]

	\renewcommand{\arraystretch}{1.3}
	\caption{Traffic  Characterization of Different Services}
	\label{Table1}
	\centering
	
	\begin{tabular}{>{\centering}p{50 pt}|>{\centering}p{35 pt}|>{\centering}p{35 pt}|p{80 pt}<{\centering}}
		\hline
		Service & Uniformity & Periodicity & Description \\
		\hline
		Video service & Weak & Strong & Concentrate in front part of the cycle \\
		\hline
		Voice service & Strong & Strong & Continuous and stable \\
		\hline
		Online chat service & Weak & Weak & Random in time and size \\
		\hline
	\end{tabular}
\end{table}

As mentioned earlier, the allocation policy in the network slicing model aims to maximize the utility function $U$ of the network, which is calculated based on the scores $\{c_i\}$ of all slices. The scores are therefore critical in determining the final training results of the algorithm. To ensure accuracy, the scores should incorporate various factors, such as average delay, transmission rate, service type, and other relevant ones. As a result, we propose the following definition given as 
\begin{equation}
	c_i=\left(r_i+\frac{l_{i,0}}{l_i}\right)^{1.1} / c_{i,0}+f_i,\label{evaluation}
\end{equation} 
where $r_i$ is the average number of requests completed in a one-time step depending on the resource allocation $\boldsymbol{k}$; $l_i$ is the average latency; $l_{i,0}$ is used to weigh the importance of latency for different slices, having the same unit with $l_i$; $c_{i,0}$ is the ideal score of slice $i$; and $f_i \in [0,1]$ 
is exclusively for the video service with weak uniformity and strong periodicity (see also Table I), and it is always zero for all other services. Particularly, $f_i$ is initialized to 0 at the start of a cycle. Once the request is processed, $f_i$ is set to 1 until the start of the next cycle.


In the MEC model, we deploy a server simulator as the ES at each NodeB, along with an agent simulator at each UE. These agent simulators generate requests of varying sizes to simulate real-world scenarios. To optimize resource utilization, we add task forwarding functionality to the server simulator. That is, if a server's processing capacity is insufficient to handle the tasks assigned to it, it will forward the tasks to the corresponding server based on instruction from the orchestrator. With this approach, the MEC system can efficiently manage the tasks and ensure that the processing load is properly distributed among the servers.

\subsection{Orchestrator}
The Orchestrator is composed of two parts: the monitor and the decider, as illustrated in Fig.~\ref{Fig2}. The monitor is implemented by Golang $1.17$, and is used to monitor the whole network and execute actions. The decider  , on the other hand, is based on Python $3.6$\cite{python} and TensorFlow $2.0.0$\cite{tensorflow2015-whitepaper}. 
For different tasks, the decider uses different algorithms: the TD3-based algorithm for network slicing and the DQN-based algorithm for MEC. Since the orchestrator only interacts with the monitor, it can be easily transplanted to different systems.


\section{Experimental Results}\label{results}
In this section, we verify the proposed TD3-based and DQN-based algorithms in both a simulation environment and the established 5G testbed. Experimental results are presented and comparisons with other allocation algorithms are drawn.

\subsection{Result of Network Slicing Model}
 During the TD3 training process, we set the learning rates of the critic network and actor network to be 0.0001 and 0.0002, respectively. The noise variance $\sigma$ is 0.1. For the first $T_1 = 40$ steps, the algorithm chooses actions randomly to explore the environment. In the course of the training process, the environments will change to simulate a dynamic real-world scenario.

We compare our TD3-based algorithm with the following algorithms:
\begin{itemize}
	\item \textbf{Optimal Allocation}: Optimal allocation  maximizes the network utility $U$ for every time step. According to Eq.~\ref{utility} and 19, it can be obtained by computing the gradient of $U$ with respect to $\boldsymbol{k}$. Note that this method does not work when the close-formed utility function is unknown.


	\item \textbf{Static Resource Allocation (SRA)}: SRA algorithm allocates the resource evenly, i.e., each slice has the same resource.
\end{itemize}

\subsubsection{Experiment in Simulation Environment} In this simplified environment, we do not consider the effect of network status.
Therefore, we set the latency $l_i$ to $0$ and ignore the parameter $f_i$ since we are studying a uniform demand for every slice. Based on Eq. 19, the score of each slice $i$ in this environment is reduced to $
     c_i=\left(r_i\right)^{1.1} / c_{i,0},$

Assume that the demand of each slice $i$ is $k_{i,d}$. Since whether the demands are satisfied is only depending on the resource allocation, we define $r_i=min\{k_i,k_{i,d}\}/k_{i,d} $, yielding the core of each slice as

\begin{equation}
     c_i=\left(\frac{min\{k_i,k_{i,d}\}}{k_{i,d}}\right)^{1.1} / c_{i,0}, \label{evaluation_simulation}
\end{equation}
where ideal score $c_{i,0}$ is given as $(c_{1,0},c_{2,0},c_{3,0})=(0.5,0.5,1)$, which reflects the possibility of satisfying the corresponding slices' demands. 

Initially, the demands $(k_{1,d},k_{2,d},k_{3,d})$ of the 3 slices are $(1,1,0.1)$.
At the mid-point, the demands of slices 1 and 2 change to $0.5$ and $1.5$, respectively. The total resource $B$ is set to $1.5$. We also have $k_{\min, i}=0.05B=0.075, k_{\max, i}=B=1.5$, for $i=1,2,3$.

Using Eq.~19 and 20, the optimal allocation is $(0.7, 0.7, 0.1)$ before the change and $(0.5, 0.9, 0.1)$ after the change. The SRA is $(0.5, 0.5, 0.5)$ consistently with even allocation.

\subsubsection{Experiment in Cloud Native wireless Network}

We conduct our experiments on an Intel i7 computer. The 5G test network and the orchestrator monitor are deployed in a virtual machine, while the decider and the server simulator are deployed in a host. Our testbed is comprised of three network slices, each with an exclusive UPF responsible for the user plane, while other NFs are shared among all slices. Each network slice has a simulated UE, and an agent simulator is deployed on each UE to simulate the services in the network slices. The agent simulators in slices 1, 2, and 3 simulate video services, voice services, and online chat services, respectively.

At the mid-point of the training, the number of services in slice 1 is reduced by half and the number of services in slice 2 increases by 50\%, leading to a change in the demands. Our system uses $B=4Mbps$ as the total bandwidth, with $k_{\min, i}=0.05B=0.2Mbps$ and $k_{\max, i}=4Mbps$ for each slice $i=1,2,3$.

In this environment, it is not possible to determine the optimal allocation in the traditional sense as a closed-form utility function $U$ for allocation $\boldsymbol{k}$ is not available. This is because the parameters $r_i$ and $l_i$ are influenced by factors beyond the allocation $\boldsymbol{k}$, such as the network and service status. Therefore, to evaluate the performance of our TD3-based algorithm, we compared it only with the SRA algorithm $(1.33Mbps, 1.33Mbps, 1.33Mbps)$.


\subsubsection{Result Analysis}

\begin{figure}[t!]
\centering
   \subfloat[]{\includegraphics[width=0.25\textwidth]{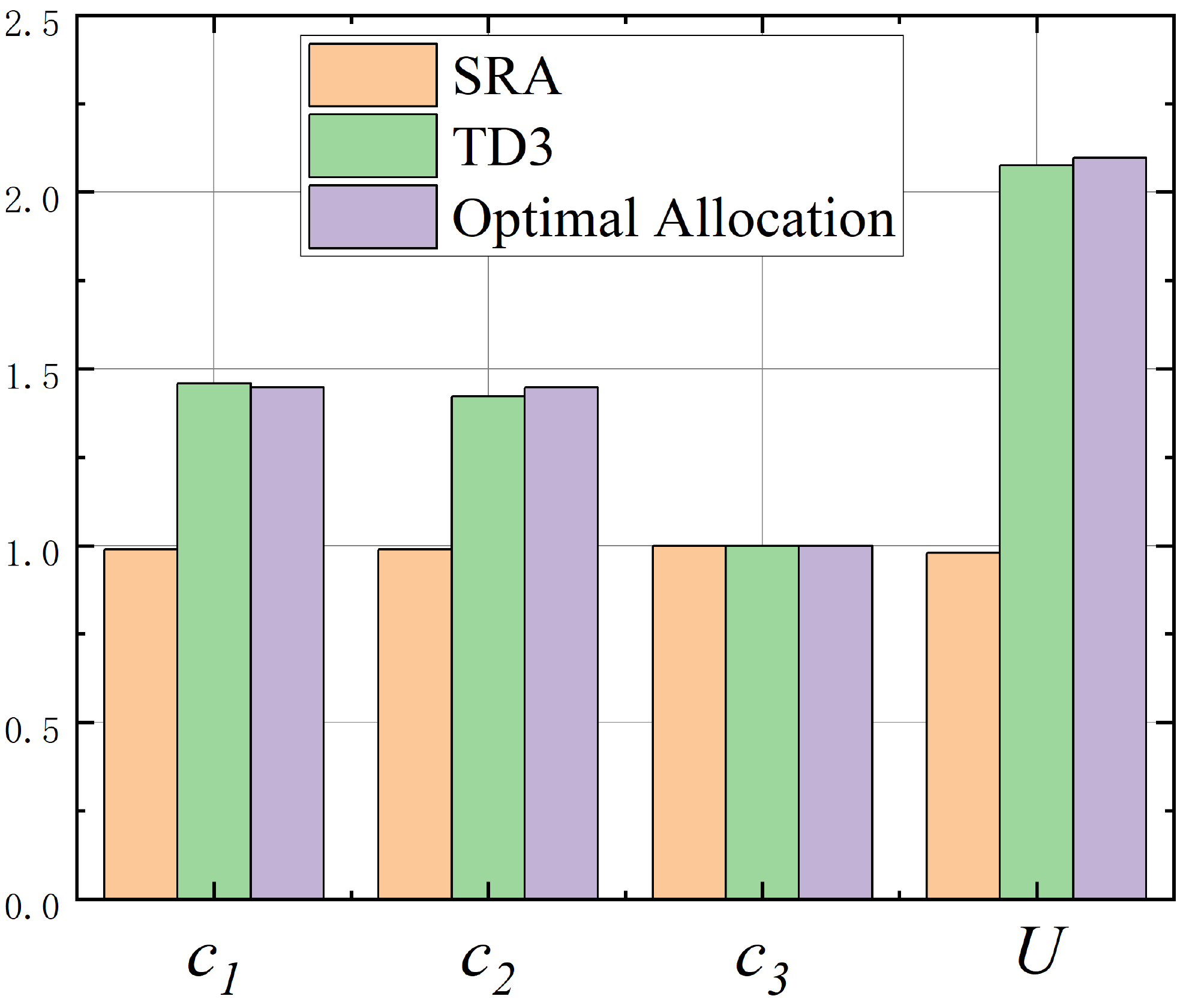}\label{Fig3}}
    \subfloat[]{\includegraphics[width=0.25\textwidth]{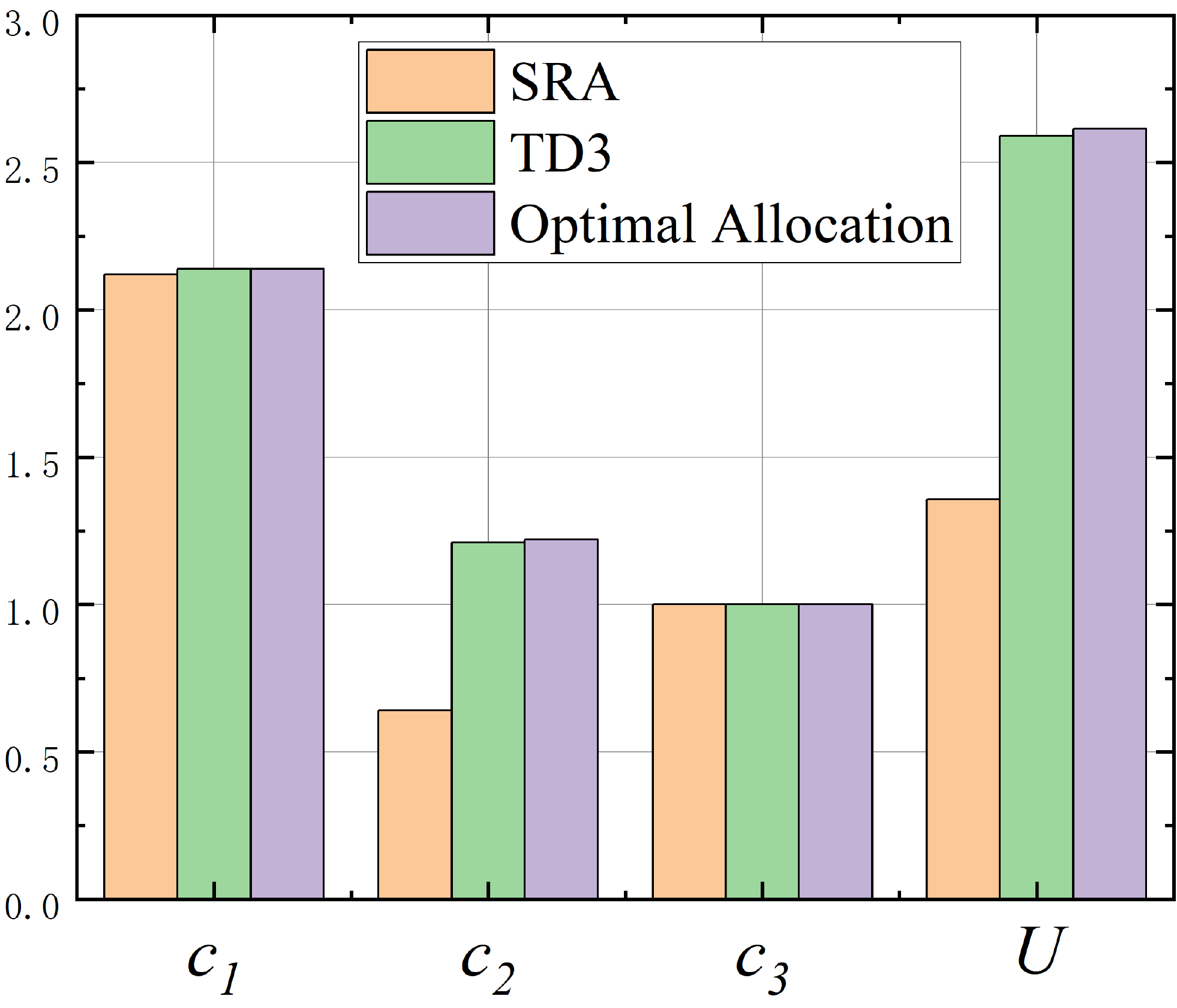}\label{Fig3b}}
    \caption{
  Comparison of the scores $\{c_1,c_2,c_3\}$ of the slices and the network's utility function $U$ in the simulation environment. (a) Before the change of the demands; (b) After the change of the demands.}
\end{figure}

For the three algorithms in the simulation environment, Fig.~\ref{Fig3} and Fig.~\ref{Fig3b} display the scores of the slices as well as the network's utility function before and after the change of the demands, separately. Since the SRA policy remains the same regardless of the demands, its performance is the worst for both cases. Specifically, for the initial demands with $(1,1,0.1)$ in Fig.~\ref{Fig3}, it is obvious that the SRA policy with $(0.5,0.5,0.5)$ allocates excessive resources for slice $3$ and insufficient resources for slice 1 and 2, resulting in only slice 3 being satisfied and having the same score as the optimal policy. In contrast, the TD3-based policy can adjust the allocation based on the demand change. Thus, as both figures illustrate, the TD3-based policy can achieve similar scores on each slice as the optimal allocation and obtain a similar network's utility function, indicating that the algorithm is effective in adopting a strategy close to optimal allocation.

\begin{figure}[t]
	\centering
    \includegraphics[height=0.35\textwidth]{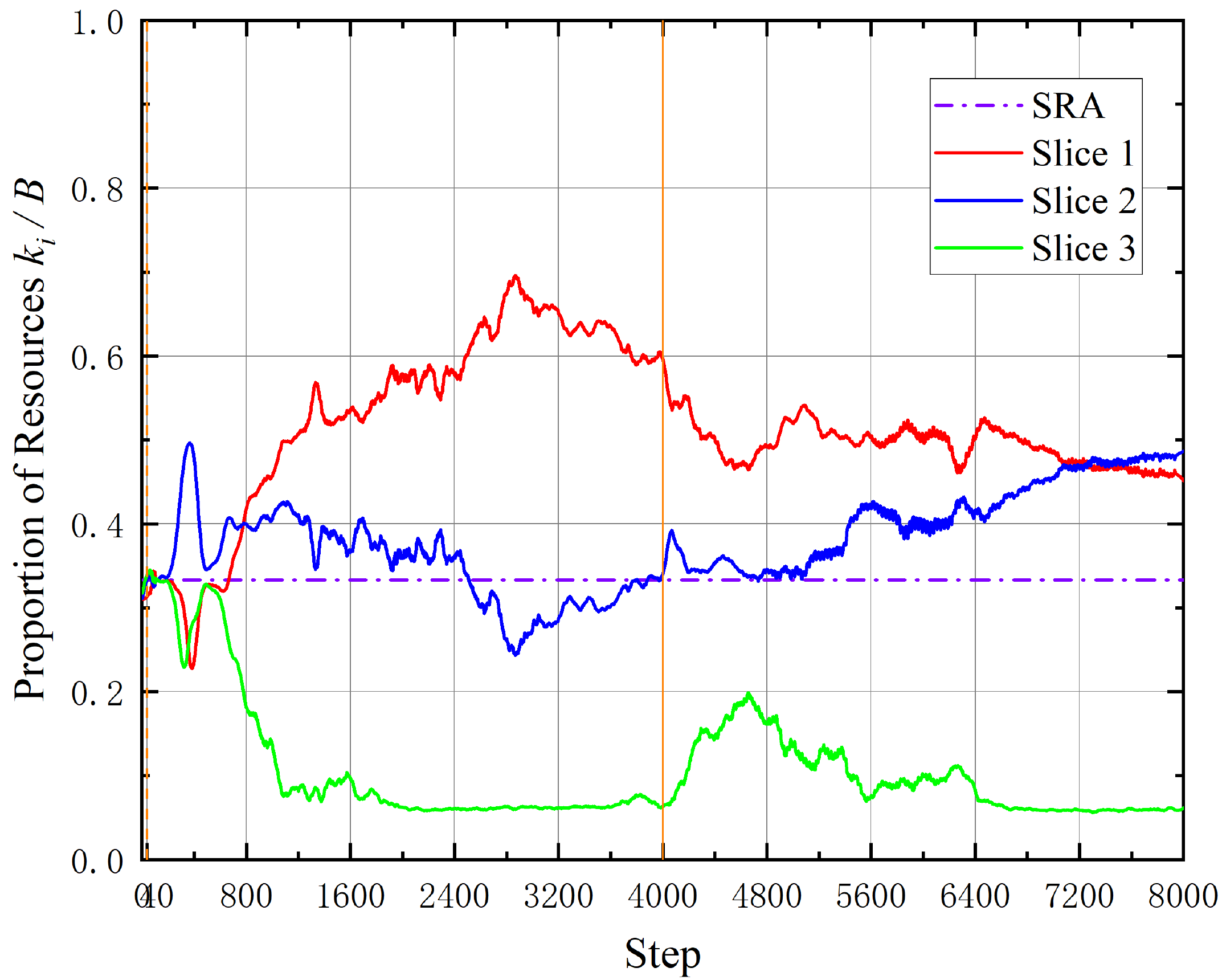}
    \caption{
    Resource proportion $k_i/B$ of each slice of the TD3-based algorithm and SRA as a function of time in the cloud native wireless network.}   \label{Fig5}
\end{figure}

\begin{figure}[t]
	\centering
    \subfloat[]{\includegraphics[width=0.25\textwidth]{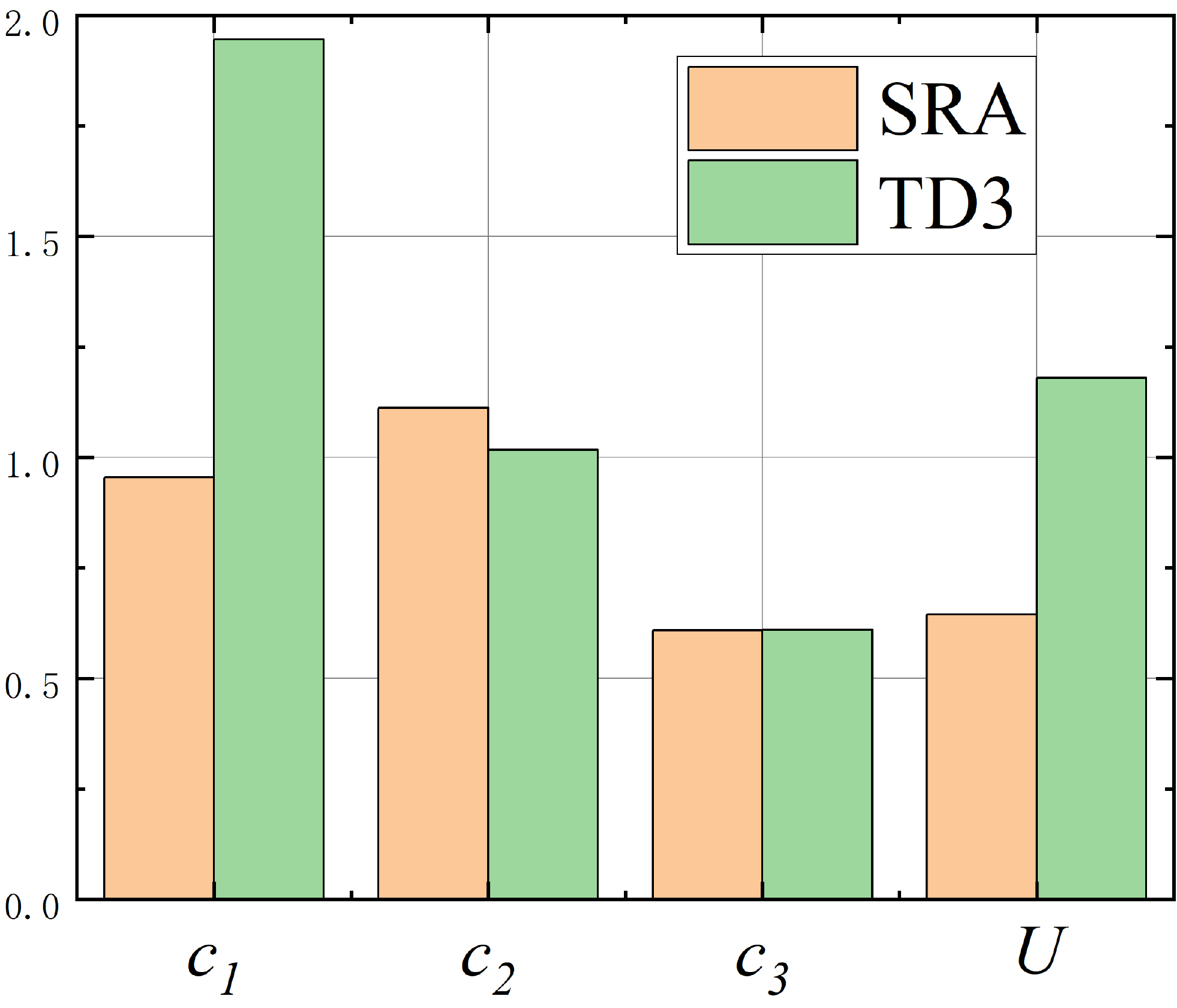}\label{Fig6}}
    \subfloat[]{\includegraphics[width=0.25\textwidth]{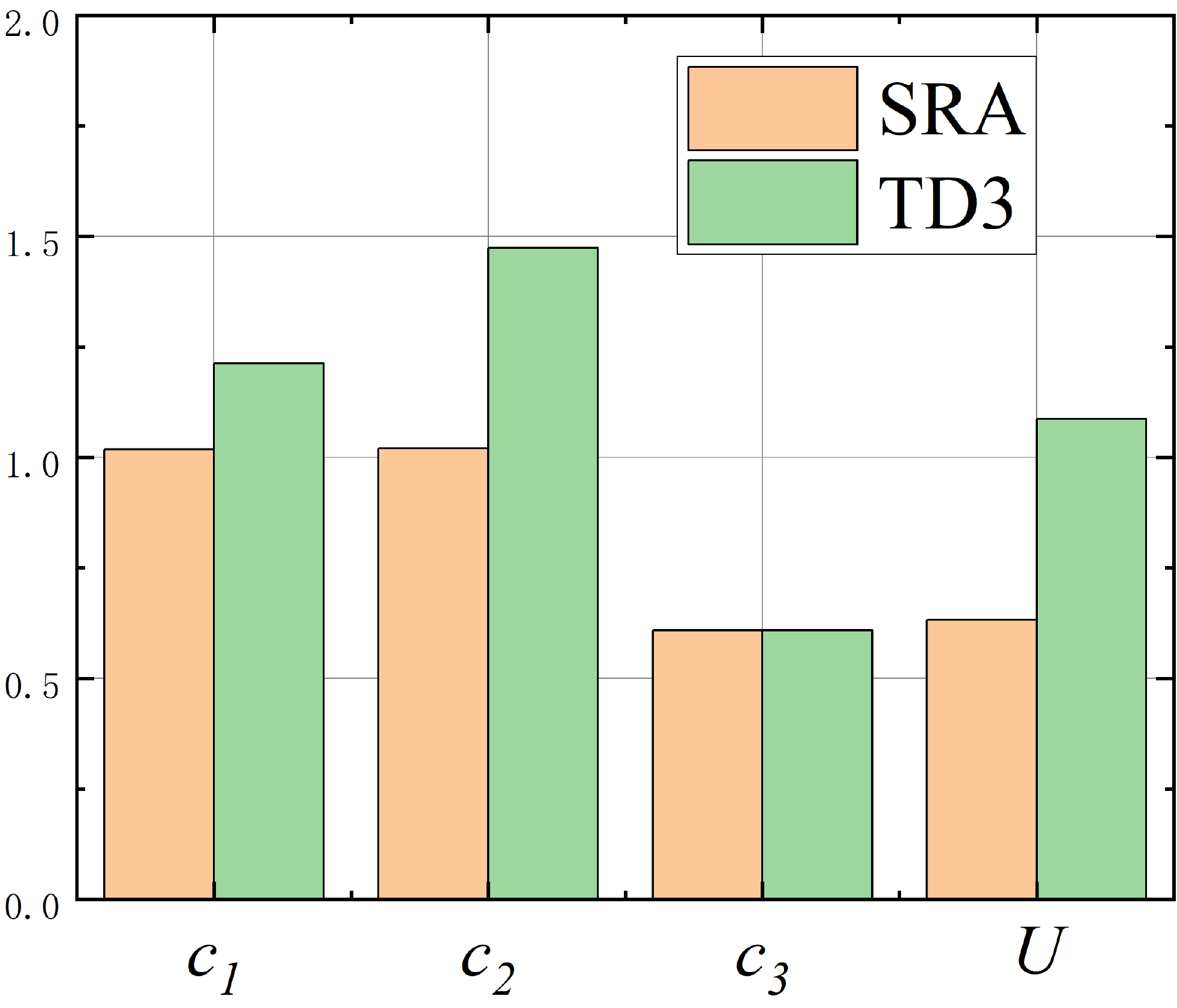}\label{Fig6v2}}
	\caption{
     Comparisons of the scores $\{c_1,c_2,c_3\}$ of the slices and network's utility function $U$ in the cloud native wireless network. (a) Before the change of the demands; (b) After the change of the demands.} 
\end{figure}

Fig.~\ref{Fig5} displays the allocation proportions $k_i/B$ of the three slices utilizing the proposed TD3-based algorithm, with the SRA strategy as a comparison. Note that SRA allocates the same amount of resources to each slice, resulting in a constant proportion of resources for each slice.

After an initial exploration of the entire action space based on a random policy during the first 40 steps, the TD3-based algorithm stabilizes its allocation results and reaches basic convergence after 3600 steps. At this point, the algorithm allocates the most resources to slice 1, followed by slice 2 and then slice 3. This allocation is reasonable given the significant differences in requirements among the network slices in our experimental environment. Slice 3 provides an online chat service, which has relatively lower demands for network resources compared to slices 1 and 2.

Following a brief stable period, the environment undergoes a change at the 4000th step, and the TD3-based algorithm correspondingly engages in new exploratory behavior within the environment. Finally, after 7600 steps, the algorithm reaches convergence. The algorithm ultimately allocates fewer resources to slice 1 and more resources to slice 2, aligning with the change in demand of the slices. The proposed algorithm's ability to automatically adjust and respond to changes in the environment is demonstrated by the result. In contrast, the SRA algorithm lacks this ability.



Fig.~\ref{Fig6} and Fig.~\ref{Fig6v2} display the scores of each slice and the network's utility function in the cloud native network before and after the environment change, respectively. The TD3-based algorithm outperforms the SRA algorithm, achieving the utility function that is 1.97 times higher before the change and 1.91 times higher after the change. The results confirm the effectiveness of the TD3-based algorithm, as well as our previous analysis, which indicates that the SRA algorithm allocates excessive resources to slice 3, leading to lower scores for other slices and the overall network. Therefore, the TD3-based algorithm is more effective and efficient, resulting in a higher network's utility function.



The TD3-based algorithm introduces some overhead compared to the SRA algorithm, which allocates resources based on a pre-defined ratio. However, this increase is acceptable as the overhead of the allocation algorithm remains stable while the network overhead increases exponentially with the number of users in each slice. In our simulation environment, the TD3-based algorithm saves approximately 18\% of the total network resources compared to the SRA algorithm while achieving the same network score. This translates to potential savings of up to 20Gbps in practical scenarios.
Therefore, the resources saved by our algorithm will far outweigh the resources it consumes, making the additional overhead acceptable.

\subsection{Result of MEC Model}

 \begin{figure}[t!]
    \centering
    \subfloat[]{\includegraphics[width=0.43\textwidth]{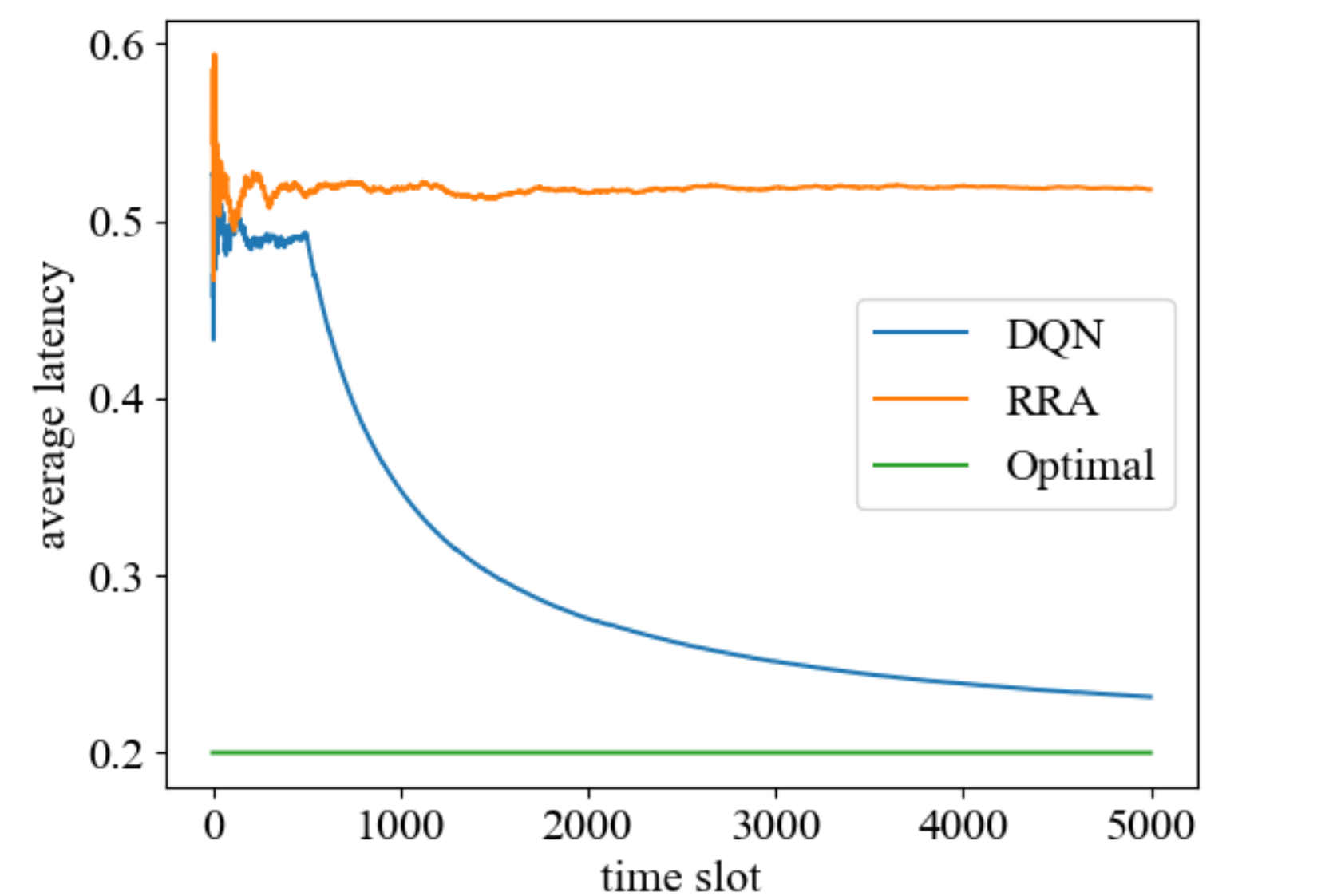}\label{Fig7}}
	\quad
    \subfloat[]{\includegraphics[width=0.43\textwidth]{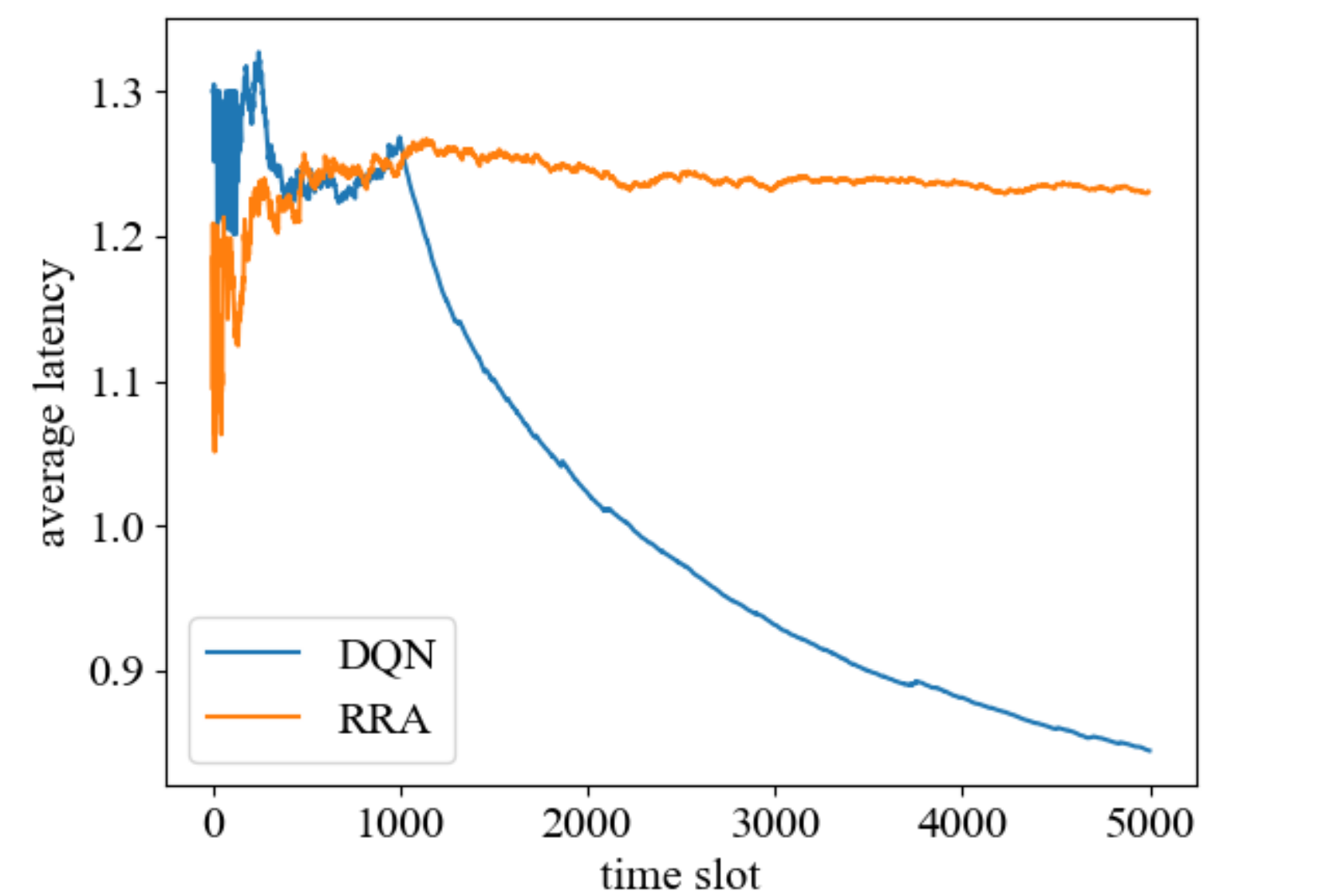}\label{Fig8}}
	\quad
    \caption{(a) Performance of the DQN-based algorithm, RRA and the optimal allocation  in simulation environment; (b) Performance of the DQN-based algorithm and RRA in cloud native 5G network.}
    \label{Fig_DQN}
\end{figure}

\subsubsection{Experiment in Simulation Environment} To validate the performance of the proposed DQN-based algorithm, we conducted tests in a simulated environment consisting of 7 ESs, including 4 eNBs and 3 gNBs. These ESs are located in two different areas, namely, Area A and Area B, with different levels of data traffic. The eNBs have smaller computing capacities than the gNBs. The initial task allocation is such that the ESs in Area A receive more tasks compared to those in Area B. The locations of these ESs are provided in Table \ref{Table 2}, and the sizes of the tasks remain constant over $T$ time slots.


\begin{table}[t!]
    \centering
    \caption{Attribute of the ESs for simulation\label{Table 2}}
    \begin{tabular}{cccccccc}
    \hline
    Number    & 1   & 2   & 3   & 4   & 5   & 6   & 7  \\ \hline
    Location  & A   & B   & A   & A   & B   & B   & A  \\
    Type      & eNB & eNB & gNB & eNB & gNB & eNB & gNB  \\ \hline
    \end{tabular}
\end{table}

We compare our proposed algorithm with another two methods, i.e., theoretically optimal strategy and random resource allocation strategy:
\begin{itemize}
	\item \textbf{Optimal Allocation}: The optimal allocation here is the allocation policy that minimizes the task processing latency for every time slot. In the simulation environment, we calculate that the minimum delay is $0.2$.
	\item \textbf{Random Resource Allocation (RRA)}: When deploying the RRA algorithm, all the ESs take random actions, meaning they transfer tasks randomly to neighboring ESs or to the Core.
\end{itemize}

In the simulation environment, $T$ is set to 5000. The first 500 time slots represent the exploration phase, during which ESs take random actions. In the exploitation stage, the decider selects an optimal allocation policy based on the information gathered from exploration, and ESs are instructed to execute the corresponding actions.

\subsubsection{Experiment in Cloud Native 5G Network} 

Using the same experimental testbed as described earlier, we set up a network with $I=7$ simulated gNodeBs, each with different computation capacities, and simulated UEs generating tasks of varying sizes at the start of each time slot. Each UE sends tasks to its corresponding gNodeB. In case an ES receives tasks that exceed its computation capacity, it transfers them to a designated ES or to the Core based on the monitor's instructions. As the resource utilization efficiency cannot be directly measured in our testbed, we use task processing latency as the optimization metric. Specifically, we aim to minimize the average latency across all processed tasks, which is used as a negative reward in the reinforcement learning process. This encourages agents to learn efficient allocation policies that can handle varying task sizes and workload distributions.

In the DQN training process, we still test the MEC setting for 5000 time slots. The first 1000 time slots are used for exploration, while the remaining ones are for exploration-exploitation where we set $\epsilon$ as $0.1$.

\begin{figure}[t!]
	\centering
	\includegraphics[width=0.88\columnwidth]{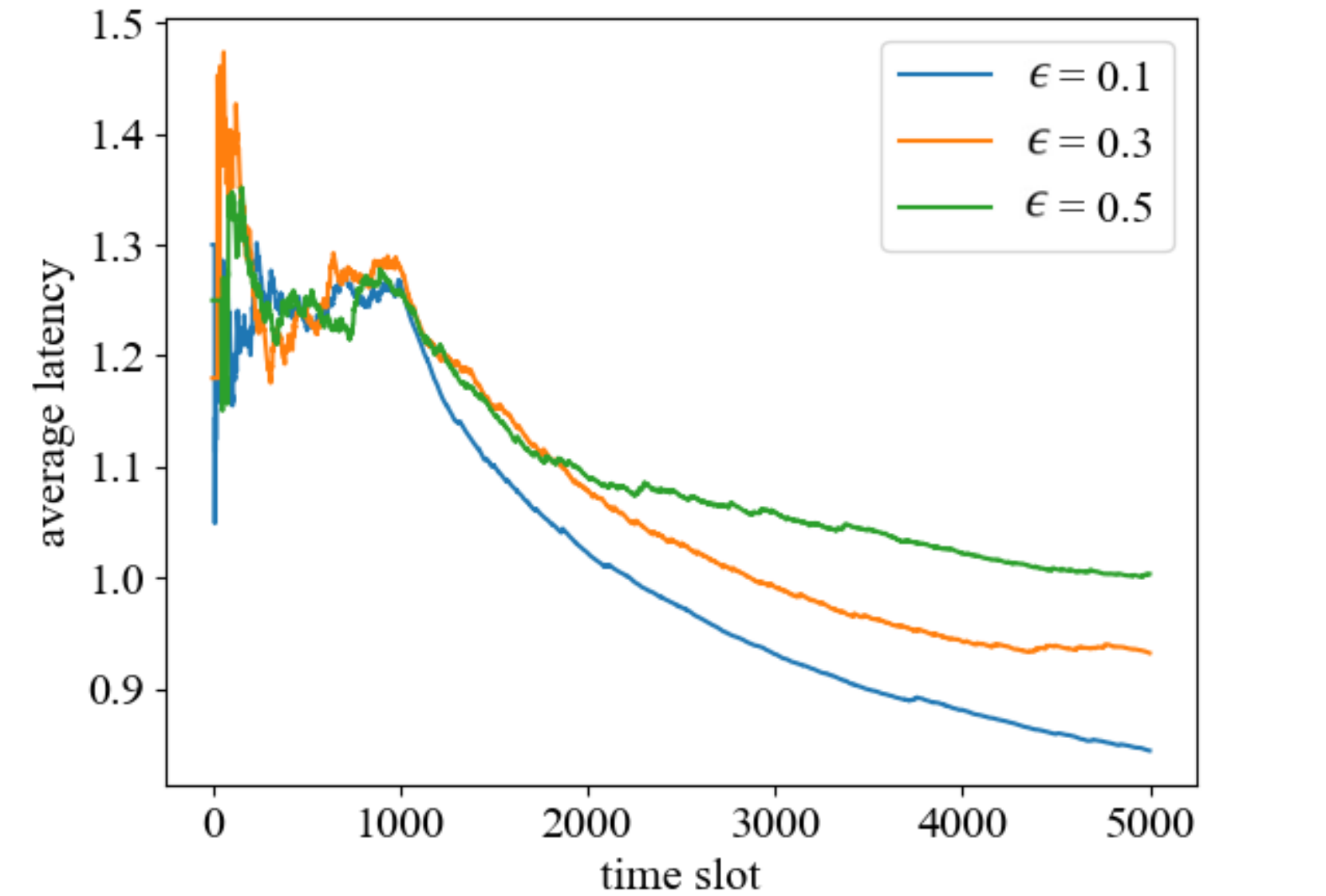}
	\caption{Performance of the DQN-based algorithm with different values of $\epsilon$.}
	\label{Fig9}
\end{figure}

\subsubsection{Result Analysis} We use the average latency to measure the performance of the DQN-based algorithm here. Fig.~\ref{Fig_DQN} shows the experiment results in a simulation environment and cloud native 5G network separately.

Fig.~\ref{Fig7} shows the performance comparison between the random policy, the DQN-based algorithm, and the optimal allocation strategy in the simulation environment. The results demonstrate that our proposed algorithm is able to significantly reduce the task processing latency compared to the random policy. Moreover, with enough time slots, the performance of the DQN-based algorithm approaches the theoretically optimal strategy.

In Fig.~\ref{Fig8}, we show the average task processing latency in the cloud native 5G network. After an initial exploration phase of 1000 time slots, the DQN-based algorithm achieves an average latency of less than 0.9, while the RRA algorithm shows no significant reduction in latency. Furthermore, our proposed DQN-based algorithm demonstrates its robustness by performing effectively with ESs having widely varying computation capacities. Although there is a slight overall increase, the average task processing latency exhibits a clear downward trend in the exploitation phase.

We also investigated the influence of $\epsilon$ by conducting additional experiments, and the results are presented in Fig.~\ref{Fig9}. We found that the smallest exploration rate, $\epsilon=0.1$, led to the shortest task processing latency under the same experimental conditions. Therefore, when using the $\epsilon$-greedy method in the exploration-exploitation phase, a smaller value of $\epsilon$ should be chosen in networks with generally stable data traffic.

\section{Conclusion and Future Work}\label{conclusion}


In this paper, we have presented our design and implementation of a cloud native wireless network, which comprises a cloud native core network based on Free5gc and a simulated RAN based on ueransim. We have addressed the resource allocation problem in two representative models, namely network slicing and MEC. To determine the optimal resource allocation strategy in both scenarios, we have proposed two algorithms based on TD3 and DQN respectively. Our experimental evaluations have demonstrated the effectiveness of our proposed algorithms in both a simulation environment and a 5G test network that we constructed. Our work highlights the significance of resource allocation in cloud native wireless networks and provides valuable insights into the application of reinforcement learning algorithms in resource allocation. 


Our work leaves open a number of future research directions. 
Firstly, it would be interesting to investigate resource allocation problems among various NFs, such as AMF and SMF, which significantly impact network performance. Secondly, exploring swarm intelligence methods with lower complexity to achieve faster convergence and combining resource allocation in the core network with that in the RAN to provide more effective solutions to the resource allocation problem in 6G would be worth studying. Moreover, as Space-Air-Ground Integrated Networks (SAGIN) are expected to play a crucial role in the 6G network, where network resources are limited and unbalanced, designing effective and robust resource allocation strategies for SAGIN is an open problem that deserves further investigation.


\ifCLASSOPTIONcaptionsoff
\newpage
\fi





%

%

\bibliographystyle{IEEEtran}
\bibliography{ref.bib}


\end{document}